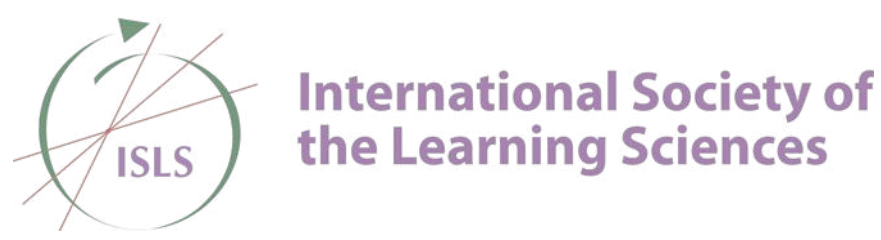

# Assessing Learning Processes with Multimodal Data in Virtual Reality Learning Environments

Eileen McGivney, Oluwatomilade Olarinde, Erica Kleinman, Fiona Shyne, Kaylah Facey, Rana Jahani, Shripad Agashe, Manav Varma, Seth Cooper, Casper Harteveld
e.mcgivney@northeastern.edu
Northeastern University

**Abstract:** Assessing learning in virtual reality (VR) environments typically relies on traditional pre-post content retention tests, revealing little about the process of learning within such immersive environments. Multimodal data from player activity in VR is promising to better measure learning processes and higher-order skills, but little research in the learning sciences has explored how such data can be combined to provide meaningful measures. To address this, we explored multimodal sources of data from a VR escape room containing hands-on logic puzzles that require problem-solving strategies and engagement in verbal metacognitive reflections. We leverage VR's affordances to immerse users in sustained dialog and promote multimodal interactions in a digital environment to understand how logfile and verbal data reveal participants' reasoning skills versus guesswork.

## Introduction

Virtual reality (VR) technologies are becoming more ubiquitous in education and training contexts as their availability increases and cost decreases. While VR faces many barriers to wide adoption including a lack of educational content, accessible design, and classroom-ready devices (Nemani, 2025), one area that has been overlooked is the challenge of assessing learning in VR (McGivney, 2024; Udeozor et al., 2023). Many studies of educational VR rely on pre-post-assessments measuring content retention or procedural skills (Hamilton et al., 2021; Wu et al., 2020). On average such studies show a slight advantage of VR for learning, but individual studies are mixed, finding positive, negative, and neutral results of VR over other media (Wu et al., 2020). One explanation is how learning is measured. For example, VR has been shown to be more consistently effective when measuring procedural skills and affective outcomes than content retention (Hamilton et al., 2021). Additionally, while learners' cognitive load and sense of presence have been studied as mediators of learning outcomes (Petersen et al., 2022; Poupard et al., 2025), little research has deeply explored the *process* of learning in VR environments. Doing so would support better learning design and improve measures of learning that align with the affordances and limitations of the technology and experiences (Engeström & Sannino, 2012).

Perhaps the most well-studied aspect of learning processes with VR has been to understand how extraneous and germane cognitive load distracts or contributes to learning in a highly stimulating environment (Mayer et al., 2022; Mulders et al., 2020). Generally, studies find heightened arousal in VR interferes with learning outcomes on post-tests and argue that distractions in VR lead to extraneous cognitive load (Makransky et al., 2019; Parong & Mayer, 2018), although some results call into question the mechanisms and directionality (Petersen et al., 2022). Research in this area has experimented with designs to mitigate cognitive load, primarily through generative learning activities like reflecting during or after a VR experience (Klingenberg et al., 2020; Parong & Mayer, 2018; Petersen et al., 2023). These studies find such activities are associated with higher learning outcomes on retention tests but typically do not track learning or reflection in-process while using the VR to understand how and why they support learning, especially of higher-order thinking or problem-solving skills.

Some studies have attempted to measure in-process learning in addition to post-VR outcomes by leveraging game play metrics. Udeozor et al. (2023) outlined a game-based assessment framework for learning in-process in VR that builds on stealth assessment (Rahimi & Shute, 2024) and provided a sample assessment of a VR laboratory health and safety game. Their sample measure used game metrics related to correct actions taken, using the proportion of correct to total moves to account for guesswork. However, it was not validated to show whether they captured learning meaningfully. Research from Johnson-Glenberg et al. (2021) illustrates how such measures in practice are ambiguous, as they assessed learning in-process of a biology game in VR and on PC with end-of-level tasks and accuracy of moves in the game. However, VR participants faced more challenges in accurately interacting with objects than those using a mouse and keyboard making the results difficult to interpret.

Despite such challenges of using in-game metrics to study learning in VR, the technology presents a wealth of other data sources that can better assess learning. For example, stealth assessments have been developed using movement and eye tracking in VR to assess psychological phenomena more accurately due to the ecological validity of behavior in VR relative to less immersive media (Chicchi Giglioli et al., 2017). VR is more immersive and gives learners more agency over the digital environment, providing affordances for learning (Makransky & Petersen, 2021) but also for assessing learning by capturing more naturalistic behavior than other media as the assessments can be even more stealthily embedded. This includes verbal data, as the heightened feeling of *being there* in the environment encourages more natural speech interactions with digital characters (Zargham et al., 2025). We are testing the benefit of such heightened verbal interaction in VR to promote deeper metacognitive reflection in a VR escape room game. We explore spoken reflections as data to assess problem-solving skills, using data from multiple modalities of movement and speech, i.e. multimodal data: logfiles and verbal reflections to assess logical reasoning performance. In this work-in-progress exploration, we ask whether measures of efficiency and performance from logfile data like proportion of right moves and time to completion can capture problem-solving strategies, and how verbal reflections can be used as data to assess problem-solving skills.

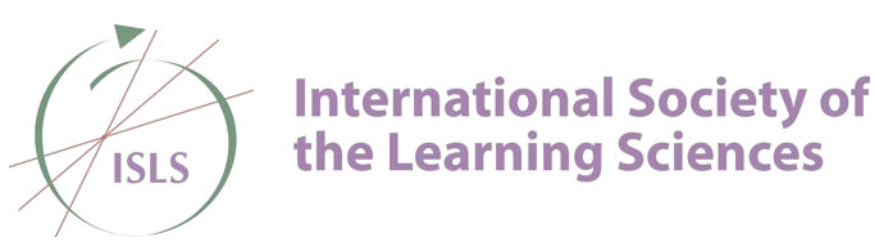

## Materials and methods

We developed a bespoke educational VR escape room-style game to study problem-solving skills via logic puzzles and instrumented it to track learner progress. The single-player VR escape room was developed using Unity, the Meta XR CORE SDK 77.0.0, and the Open XR Plugin 1.14.0. Participants explore an abandoned house where they can access a kitchen and a greenhouse with five logic puzzles related to cooking and gardening that the participant must solve following a hub-and-spoke style: They solve four simpler puzzles with simple logical problem-solving insights and unlock clues to the culminating "hub" puzzle that requiring more advanced logical insights. The spoke puzzles use hands-on interaction with game objects (watering plants, pouring sauce onto pasta), while the hub puzzle uses a more traditional grid interaction. Interaction is via hand tracking, not controllers, including gestures for teleporting beyond moving in the physical space allowed. Figure 1 depicts the layout of the escape room, and a screen capture of a participant solving the hub puzzle that illustrates how they filled a grid with correct answers (green pins) and eliminated options (red pins) to create a soup recipe. Because the broader project studies metacognitive reflection in learning with VR, learners were prompted to reflect on their process while solving the puzzles. Prompting occurred based on standardized triggers for all participants based on best practices of escape room game masters (Kleinman et al., 2025): 30 seconds idling, taking no actions, attempting to use objects unrelated to the puzzle to solve it, or expressing frustration or confusion. Prompts were issued verbally by a research team member present in the physical room observing the activity in VR, and drawn from a prompt bank from prior research (Kleinman et al., 2022; Salehi, 2018) including: *What assumptions are you making about the puzzle, and how can you test those assumptions?* and *What strategies did you use in prior puzzles that were effective? How can you apply them here?* Participants were asked to respond verbally in one or two complete sentences. The purpose of reflective prompting was to help learners process deeper structures of logical problem solving and develop strategies, rather than focusing on the mechanics of the VR environment.

### Participants

A convenience sample of 14 people was recruited. The mean age was 25 (range: 19-33). Nine identified as female, four as male, and one as non-binary. Six had no experience with escape rooms, five identified as novices, and three as intermediate users. Participants' prior VR use: frequent (1), a few times (12), and never (1). Participants consented to the study and received a $25 gift card. This study was approved by the Northeastern University IRB.

**Figure 1**
*Escape Room Puzzle Layout (Left) and Hub Puzzle Detail (Right).*

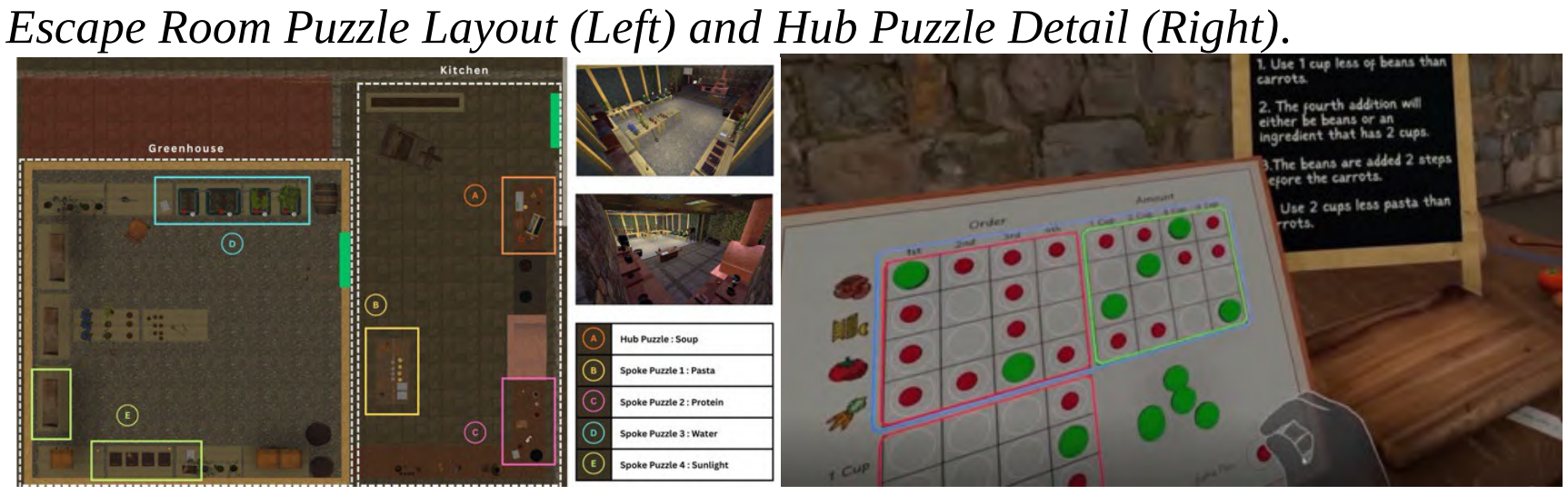

### Data and analysis

*Logfile data on participants' moves.* Each interaction a participant had with an object in the VR environment was logged by the system, and in this exploration we use moves only on the hub puzzle to assess logical problem solving strategies, as solving the easier spoke puzzles helps participants learn both the mechanics of the escape room game and logical reasoning strategies to efficiently solve the more complex hub puzzle. Each move was interpreted by a puzzle solving algorithm ("the solver") developed by research team members (Shyne et al., 2024). It identifies correct moves and the potential logical reasoning used by consulting a list of possible moves based on the board state and puzzle hints. For example, the solver intuits the participant eliminated an option because the clues available to them indicated that ingredient could not be first in the recipe based on its relation to another ingredient. This logfile data was used to calculate the number of moves a participant used to solve the hub puzzle, the number of those moves that were correct, and the amount of time it took them to solve the puzzle. For correct moves that are not in the solver's list of possible moves, the solver marks the reasoning as "unknown," meaning the participant could have used logical reasoning, but is not apparent based on the current state of the board.

*Video recordings for participant insight and guesswork.* For each participant's hub puzzle play, we manually coded each correct move based on the participant's video and audio to validate whether the participant was evidently using logical reasoning insights as suggested by the solver, using guessing/trial-and-error, using some form of strategy that was not aligned with the logical reasoning insight, or it was unclear what led them to that move. Evidence they used logical reasoning insights aligned with the solver came from statements about how they were using the clues to determine or eliminate options, such as *"So I can eliminate the carrot, [it] won't be 1st or 2nd because yeah, [beans is] 2 cups before"*(P6). Guessing was identified by moves on the hub puzzle before unlocking its hints, such as *"I'm assuming that says beans,"*(P8) or verbalizing guessing as strategy: *"I'm just going to try and guess something and hope that works and if it doesn't, I'll try something else"*(P2). We summed the number of correct moves based on insight and guessing to estimate proportions of total moves for each participant. Moves not classified as insight-based or guessing were not included in the analysis.

*Audio recordings for depth of reflection.* To assess quality and depth of metacognitive reflection, two authors conducted deductive coding on transcripts of verbal reflections (*Cohen's Kappa = .80*). The codebook was adapted from the situational awareness model, appropriate for assessing verbal reflections conducted during interaction with a system rather than post-activity reflections, and has been associated with metacognitive processing in digital environments and cognitive load in VR (Mohaddesi et al., 2023; Tomori & Ogunseiju, 2025). Each response to a prompt was coded as perception (recognizing relevant elements), comprehension (understanding the goal), or projection (project the future by describing key steps to solve the puzzle), which indicate increasing depth. These were assigned a score of 1 (perception), 2 (comprehension), or 3 (projection), summed and averaged to give a score for the mean depth of each participant's reflection responses.

**Table 1**
*Participant Scores Across Game Play Metrics, Move Coding, and Reflection Coding.*

| | Game Play Metrics – Hub Puzzle | | | Manual Move Coding- Proportion Correct | | Reflection Response Depth (Count) | | | | |
|---|---|---|---|---|---|---|---|---|---|---|
| Participant Number | Completion Time | Total Moves | Proportion Correct of Total Moves | With Insight of Correct Moves | Guessing | Reflection Prompts | Perception (1) | Comprehension (2) | Projection (3) | Mean Reflection Depth |
| **Mean** | **8:31** | **43** | **51%** | **58%** | **10%** | **7** | **3** | **4** | **1** | **1.73** |
| 2 | 4:31 | 35 | 54% | 42% | 16% | 6 | 2 | 4 | 0 | 1.67 |
| 3 | 5:43 | 26 | 50% | 31% | 38% | 8 | 4 | 4 | 0 | 1.5 |
| 4 | 5:47 | 16 | 50% | 0% | 0% | 6 | 5 | 1 | 0 | 1.17 |
| 5 | 10:01 | 66 | 52% | 59% | 0% | 9 | 5 | 3 | 1 | 1.56 |
| 6 | 9:53 | 29 | 72% | 81% | 5% | 10 | 2 | 6 | 2 | 2 |
| 7 | 10:44 | 95 | 23% | 55% | 5% | 4 | 2 | 1 | 1 | 1.75 |
| 8 | 1:38 | 12 | 75% | 56% | 33% | 6 | 1 | 5 | 0 | 1.83 |
| 12 | 2:50 | 24 | 38% | 67% | 0% | 6 | 3 | 2 | 1 | 1.67 |
| 14 | 12:02 | 36 | 92% | 88% | 0% | 10 | 1 | 8 | 1 | 2 |
| 16 | 6:34 | 40 | 33% | 38% | 38% | 7 | 3 | 4 | 0 | 1.57 |
| 20 | 12:31 | 37 | 35% | 54% | 0% | 9 | 5 | 4 | 0 | 1.44 |
| 21 | 10:19 | 45 | 64% | 90% | 0% | 3 | 1 | 2 | 0 | 1.67 |
| 22 | 15:06 | 100 | 32% | 75% | 0% | 7 | 1 | 2 | 4 | 2.42 |
| 23 | 11:21 | 39 | 46% | 78% | 0% | 8 | 2 | 4 | 2 | 2 |

## Results

Table 1 displays each participant's scores. Values better than the mean are highlighted in green (*less* time and moves for completion; *more* proportion correct, insight, and reflection), and those worse than the mean in red.

Efficiency measures are not good metrics for problem-solving skills.

Because they played four simpler logic puzzles before attempting the hub puzzle, we anticipated participants who learned the most about logical reasoning would solve the hub puzzle more efficiently. The participants who solved the hub puzzle in below-mean time and with below-mean number of moves were 2, 3, 4, 8, 12, 16, each solved the puzzle less than 8:30 and 43 moves. However, using verbal data reveals they were not leveraging as many logical reasoning skills as participants who took more time and moves. Specifically, participants 3, 8, and 16 earned more than a third of their correct moves through guessing and were below the mean proportion of correct moves that could be attributed to logical insights. We also found a negative correlation between completion time and correct moves attributed to guessing, at a nearly statistically significant level ($r$= -.53, $t(12)$ = -2.17, $p$=.051).

Verbal reflection data is necessary to understand participant performance.

Manual coding assessed the match between the solver's logical reasoning insights needed to make a correct move and the participant's verbal reflections, revealing instances of guessing despite having the information needed to make a correct move. For example, P3 engaged in a high proportion of guessing to achieve correct moves. On their $19^{th}$ move in the hub puzzle, they placed the green marker for pasta on the board to indicate it is 2 cups. The solver intuited from this that the participant was using the hint that "Pasta is two cups less than carrots," which would require the logical reasoning that since carrots are 4 cups, they can apply the rest of the hint ("*apply before two spots*"). However, as the participant speaks about their process they say, "I'm just going to give it a guess and do that…let's see how that plays out." This is different from their approach on their $22^{nd}$ move, which they place the green marker correctly and stating "…$4^{th}$ ingredient will either be beans or an ingredient that has 2 cups – and 2 cups is pasta," using a transitive logical insight mapping the hint to values of both amount and order. Without spoken data it is not possible to verify when a participant is using the reasoning their solving behavior suggests.

Greater proportions of correct moves and logical reasoning are associated with depth of reflection.

The number of reflections was not associated with characteristics of participant performance such as number of correct moves or correct moves using logical reasoning insights: reflecting more often was not associated with better performance on the hub puzzle. However, there was an association between depth reflections and use of logical reasoning on the hub puzzle. A higher average reflection depth (mean rating on 1-3 scale from perception to projection), was correlated to total number of correct moves made at a nearly statistically significant level ($r$ = .53, $t(12)$ = 2.16, $p$=.052). Reflection depth and proportion of correct moves using logical reasoning were correlated ($r$=.73, $t(12)$ = 3.73, $p$=.002): Reflections beyond perceiving the environment were more likely to use logical reasoning. This is not due to participants simply being more vocal making their reasoning easier to detect.

## Conclusion

While the promise of VR for learning has long been understood as providing deeply situated learning opportunities for complex skills, but to realize this promise better measures of the process, not just outcomes, of learning are needed. VR captures a wealth of motion and biometric data, but increased immersion also presents opportunities to capture data from spoken interactions that feel natural in VR. This work-in-progress study explored the potential for such verbal data to support assessing learners' process in VR, revealing its benefits to identify logical reasoning versus guessing. Next steps include developing an AI character to conduct reflective prompting, assess the difference in amount and depth of spoken reflection in VR with this character relative to a PC version, and incorporate automated analysis of verbal data into the solver's understanding of learners' logical reasoning. Evidence here informs that automated analysis and future generalizable measures for VR learning environments.

## Acknowledgments

This work is funded by NSF grant #2418612, *AI-Empowered Reflective Learning Across the Virtuality Spectrum.*